\documentclass{amsart}
\usepackage{graphicx}
\usepackage[vflt]{floatflt}
\usepackage[utf8]{inputenc}
\usepackage[english]{babel}
\usepackage{color}
\usepackage{fancyvrb}

\definecolor{red}{rgb}{1,0,0}
\definecolor{black}{rgb}{0,0,0}
\definecolor{light-gray-one}{gray}{0.85}
\definecolor{light-gray-two}{gray}{0.65}

\def \CC {{\mathbb{C}}}

\let\oldmarginpar\marginpar
\renewcommand\marginpar[1]{\-\oldmarginpar[\raggedleft\footnotesize #1]
{\raggedright\footnotesize #1}}

\begin{document}
\author{Alessandro Rosa}
\title[Kleinian groups - Index search]{A new indexed approach to render the attractors of Kleinian groups}
\date{\today}
\email{alessandro.a.rosa@gmail.com} \maketitle \vspace{-1.0cm}

\begin{abstract}
One widespread procedure to render the attractor of Kleinian
groups, published in the renown book \cite{Indras-2002} and based
upon a combinatorial tree model, wants huge memory resources to
compute and store all the words required. We will present here a
new faster and lighter version which drops the original words
array and pulls out words from integer numbers.
\end{abstract}

\section{Introduction: some definitions}\label{introduction}
Let $K$ be a group of one-to-one relations. One model binds the
elements of $K$ to strings of letters, because these symbols show
up in two cases, distinguishing the elements from their inverses:
`$a$' (lower case) and `$A$' (upper case).

The \emph{generating set} $G$ is the smallest subgroup of $K$ such
that elements of $K$ are expressed as the combination, termed
\emph{multiplication}, of elements of $G$, often tagged with
single letters,\footnote{For example $a\circ b$, but the operator
$\circ$ is often omitted for sake of brevity.} and collecting into
the \emph{alphabet} of $K$. Multiplication corresponds, lexically,
to concatenation of letters into one string: the so-called
\emph{word}. Words resemble to algorithms, enjoying symbolic
(code) and operative (run) features. There are finite
($bbbbaBAbA$) or infinite\footnote{The overline symbol marks the
period, like for numbers.} ($\overline{b}bbbaBAbA$) words and the
reading order, left-to-right ($LR$) or right-to-left ($RL$),
drives the letters/generators picking. Let $W=abA$ in RL, then
$z_1 = A(z_0), z_2=b(z_1), z_3 = a(z_2)$ returns a sequence of
values $z_n$, the \emph{orbit}. The last orbit element is defined
here as \emph{word value}. Any subword, returning the identity map
$I$, is said \emph{crash word}, provoking the \emph{cancellation}
of letters and returning a \emph{reduced} word. Let $W=aBbA$: we
have two cancellations, $bB$ and $aA$, and $W$ reduces to $I$.

Words in $K$ converge uniformly to limit cycles,\footnote{Every
$K$ is a \emph{convergence group}; see \cite{GeMa-1994}, pp.
334--340.\label{footnote_c_convergence}} collectively defined as
the \emph{attractor}. Generators and words show up in a twofold
(lexical and geometric) nature: as symbol/point and as
concatenation/orbit respectively. Such a duality extends to
groups, in terms of words/attractor.\footnote{Alternatively
defined the `limit set'.}

\section{Basic setup}\label{basicsetup}
\begin{figure}[htb]
\begin{tabular}{c}
\input{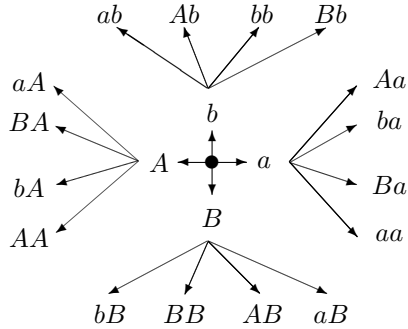}
\end{tabular}
\caption{\textbf{Original tree}. Enumeration of all possible
concatenations of symbols up to depth $2$.}\label{tree_full_01}
\end{figure}

Working with attractors wants a sufficient degree of freedom and
to consider all words in the group. So we step back to the
abstraction of strings and symbols, because we need a `malleable'
setup to work with: \emph{any concatenation of letters up to
bounded length} $d$. According to the theory of enumerative
combinatorics, it is graphically feasible through a $m$-branched
tree (see fig. \ref{tree_full_01}), where $m$ is the alphabet
length.
Luckily, the theory of combinatorial groups\footnote{Pioneered by
Sir Arthur Cayley during 1850s. Refer to \cite{Cayley-1854}.} can
set up close links between tree graphs and groups: generators and
words interact with the concepts of \emph{node}, \emph{path},
\emph{depth}, \emph{root}, \emph{leaf}, \emph{parent},
\emph{child}. A comfortable tool to condense the rules of group
generation is the \emph{presentation}. We account two versions:
the \emph{Cayley multiplication table}, for \emph{finitely
generated groups},\footnote{Equipped with a finite number of
generators and of relations between them.} including all
multiplicative combinations between elements of $G$; and the
so-called \emph{group presentation}, a compact list of generators,
the \emph{relators} $R$, and \emph{relations} $S$
\cite{LynSch-1977}, often of crash kind: $\langle R | S
\rangle$.\footnote{Group presentations can be considered as a
generalization of Cayley tables.} When supported by a
presentation, the tree shows as the easiest graphical expedient to
explain the group generation. Figure \ref{tree_full_01} shows the
\emph{original tree}, not related to group presentation. We will
work with trees of bounded depth $d<+\infty$.

\begin{figure}[htb]
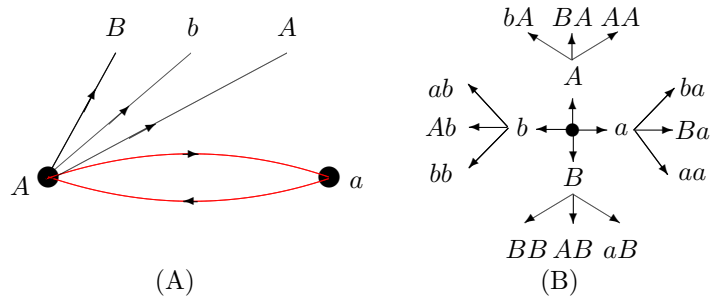

\begin{tabular}{ccc}
  \input{crashword.pic}\label{pix_crashword}
  & &
  \input{tree.pruned.alphabetic.pic}\label{pix_tree_alpha}\\
  (A) & & (B)\\
\end{tabular}
\caption{\textbf{Pruning}. (A) New nodes are pruned if
producing a crash word (in red). (B) The pruned
tree.}\label{tree_tabular_01}
\end{figure}

\section{Once-punctured torus groups}\label{oncepuncturedtorusgroups}
Let $K$ be a Kleinian group, a discrete group of orientation
preserving conformal maps $M$. In recent times, this topic gained
more interest from the popular audience as it was dragged by the
caravan of fractals, due to the close links to Julia
sets.\footnote{Gaston Julia was the first to set this analogy in
1918, while studying the iterations of functions in one complex
variable. Refer to \cite{Air-2011}.} Let $M$ be a linear
fractional map $(mz+n)/(pz+q)$ in one complex variable $z\in\CC$.
We are interested into the quasi-Fuchsian subfamily of $K$ and we
will work with $4$ generators $a,b,A,B$: the topological model is
the once-punctured torus and it shows as the free product $K = G *
H$, $G=\{a,A\}, H=\{b,B\}$. The presentation is $\langle x,X |
xX=I \rangle$ or, since Kleinian groups are finitely generated,
the Cayley table \ref{T11CayleyTable}. $K$ is free because no more
relations besides trivial ones are listed. We will discuss the
role of Cayley table later in section \ref{CMT}. This presentation
prunes the original tree in fig. \ref{tree_full_01} from nodes
related to strings with crash words \emph{Aa, aA, Bb, bB} (fig.
\ref{tree_tabular_01}/A at p. \pageref{tree_tabular_01}) that send
points forth and back like in a cycle: $z_1 = A(z_0), z_0=a(z_1)$.
Here the identity map halts the branching action of the tree and
results \emph{only when} the next child node has same letter as of
its parent, but inverse case. The goal is to have no reduced words
(which basically duplicate other nodes) and we get the pruned tree
in fig. \ref{tree_tabular_01}/B.

\begin{table}[h]
  \centering
\begin{tabular}{|c|c|c|c|c|c|}
  \hline
             & \textbf{I} & \textbf{a} & \textbf{b} & \textbf{A} & \textbf{B} \\
  \hline
  \textbf{I} & $I$ & $a$ & $b$ & $A$ & $B$ \\
  \hline
  \textbf{a} & $a$ & $a$ & $b$ & $I$ & $B$ \\
  \hline
  \textbf{b} & $b$ & $a$ & $b$ & $A$ & $I$ \\
  \hline
  \textbf{A} & $A$ & $I$ & $b$ & $A$ & $B$ \\
  \hline
  \textbf{B} & $B$ & $a$ & $I$ & $A$ & $B$ \\
  \hline
\end{tabular}
\vspace{0.4cm} \caption{\textbf{Cayley multiplication table for
once-punctured torus groups.}}\label{T11CayleyTable}
\end{table}

\section{The revised deterministic approach}\label{reviseddeterministic}
The problem of rendering the attractor\footnote{The renderings of
attractors in this article have been computed through author's
software `Circles':
http://alessandrorosa.altervista.org/circles/\label{footnote_circle}}
of $K$ has been studied thoroughly, in terms of \emph{automatic
groups},\footnote{Any finitely generated group equipped with a
finite state automata. Refer to \cite{ECHLPT-1992}, p. 356.} only
in \cite{McPaRe-1994,Indras-2002}, as far as the author knows.
\begin{figure}[h]
  \begin{tabular}{c}
  \includegraphics[height=3cm]{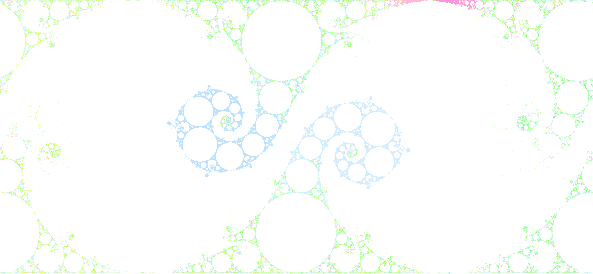}
  \end{tabular}
  \vspace{-0.3cm}
  \caption{\textbf{Probabilistic rendering.} Attractor for the parameter $\mu=-0.097+1.838i$ in
  the Maskit T1,1 embedding, rendered via a `boosted-up' modification.
  Only 1.048.576 words has been required to enhance sharp details.}
  \label{probabilisticpix}
\end{figure}

\begin{figure}[htb]
  \begin{tabular}{c}
  \includegraphics[height=3cm]{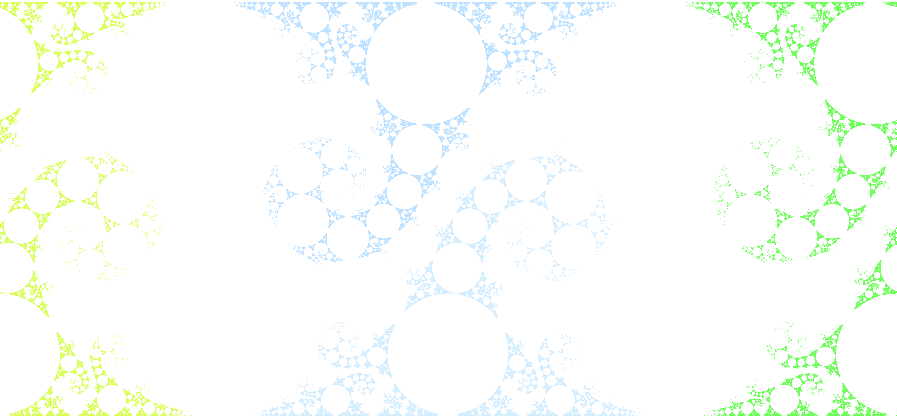}
  \end{tabular}
  \vspace{-0.3cm}
  \caption{\textbf{Index-search approach.} Same attractor as in picture \ref{probabilisticpix},
  rendered through a tree of depth $14$, counting about $350$ millions of words. In terms of
  original deterministic approach, the dictionary would weight more then $4$ Gigabytes.}
  \label{pix01}
\end{figure}

Two approaches have been developed. One is \emph{probabilistic},
not relying on tree model and working on one only word/orbit,
which gets longer and longer as generators are appended through
random picks, given a probability law.\footnote{A sketch is
available in \cite{Indras-2002}, p. 152. This algorithm can be
boosted up via commutators, similarly to a technique for the
deterministic approach, discussed in \cite{Indras-2002}, pp. 181,
248.} The second is \emph{deterministic}, where words obey to the
combinatorial tree model. The larger the depth, the longer the
word, the finer the rendering: it is the essential principle of
this approach, needing to run millions of words for sharp
results.\footnote{Authors of \cite{Indras-2002} had to pull
additional manipulations (so-called `repetends') out of the hat,
to catch up more details, because orbits run slower and slower in
the neighborhoods of (nearly) parabolic points.} The original
implementation casts a huge array of words - the
\emph{dictionary},\footnote{See p. 114 of \cite{Indras-2002}
suggested cardinality $10E7$ or more, refer to caption of fig.
\ref{pix01}.} demanding a very expensive run, for fine quality
results, in terms of memory allocation and addressing. The core of
the algorithm is essentially the same as of the original one, so
it is not intended to return finer quality renderings.

\begin{figure}[h]
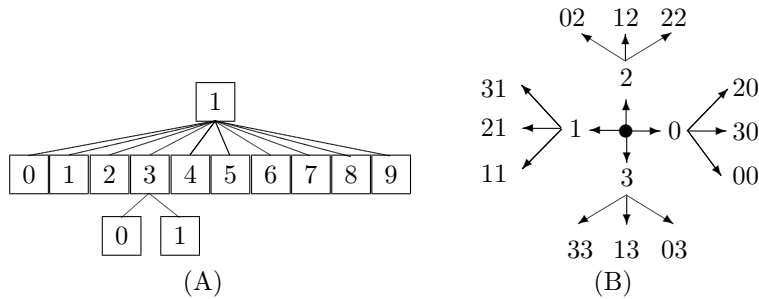

\begin{tabular}{ccc}
  \input{digitbox01.pic}\label{pix_digit_box}
  & &
  \input{tree.digit.pic}\label{pix_tree_digit}\\
  (A) & & (B)
\end{tabular}
\caption{\textbf{First stage}. (A) Numerical paths: the mid row
shows numbers composition from 10 to 19. The
bottom row shows one next step. (B) Digital version
of the lexical pruned tree.}\label{tree_tabular_02}
\end{figure}

We are going to explain a two-stages strategy revisiting the
pruned tree of letters. Numbers are the synthesis of two entities:
\emph{symbol} and \emph{value}. Consider the set $\{10, 11, 12,
\dots, 19\}$ in terms of symbols: the elements come from appending
one digit on the right of `$1$', like a new branch of a tree (see
fig. \ref{tree_tabular_02}/A). \emph{It makes sense to review
numbers in terms of paths.} Given the mapping $[a\rightarrow 0,
A\rightarrow 1, b\rightarrow 2, B\rightarrow 3]$, we replace
letters with digits in the tree at p. \pageref{pix_tree_alpha}.
Starting from the root at depth 0, corresponding to $I$ (empty
string), we move to the left branch at depth $1$ and get the
\emph{subword} `$1$'; we route to one of next branches and get
`$31$' at depth $2$. Each node binds to a unique path, to one only
chain of digits and we pull out the \emph{digital tree} in fig.
\ref{tree_tabular_02}/B.\footnote{The concept of an indexed tree
is widely used in Computer Science, from disk storage
representation to frequencies and data compression
\cite{Fenwick-1994}.}

\begin{figure}[htb]
\input{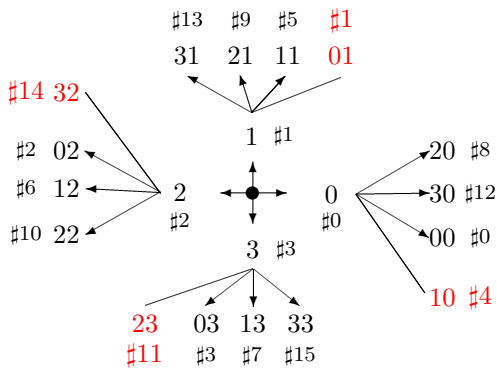}
\caption{\textbf{Second stage}. Indexed version of the digital
tree: red nodes mark crash words.}\vspace{-0.2cm}
\label{tree_indexed_pic}
\end{figure}

The master plan is to move from symbols to digits and finally to
indexes (numbers). Now notice that words include digits from $0$
to $m-1$ (here $m=4$) and nodes bind to base-$m$ numbers, turning
into indexes in base-10 (prefixed by `$\sharp$' in fig.
\ref{tree_indexed_pic}). Indexes count the appearing order of
nodes, during the whole tree growth. The notation
$31_{4\leftarrow}$ indicates that $31$ is written in base-$4$ and
read in RL order: it amounts to $4_{10\leftarrow}$. For example,
$Ba\Rightarrow 21_{4\leftarrow}\Rightarrow 5_{10\leftarrow}$. We
finally get the \emph{indexed tree} in fig.
\ref{tree_indexed_pic}/A), whose nodes bind to base-$10$ indexes.

So \emph{we do not need to store the words: everything we want is
already `coded' inside integer numbers. We have just to count
them!}

\section{Cayley Multiplication Tables}\label{CMT}
Cayley tables portray all multiplicative combinations between
pairs of elements of $G$ and are very intuitive for code
implementation: they are homologue to \emph{state transition
tables} and can be supported by a \emph{finite state automaton}
(FSA) to compute word runs \cite{HoUl-1979}. In terms of values,
words become dynamical systems whose `states' match to cells
value. Cayley tables prune the original tree, depending on whether
the final state is of crash kind (indexed with $0$) or not. How ?
\begin{table}[h]
\centering
\begin{minipage}{0.3\textwidth}
\vspace{-0.3cm}
\begin{center}
\begin{tabular}{|c|c|c|c|c|c|}
  \hline
      & $0$ & $1$ & $2$ & $3$ & $4$ \\
  \hline
  $0$ & $0$ & $1$ & $2$ & $3$ & $4$ \\
  \hline
  $1$ & $1$ & $1$ & $2$ & $0$ & $4$ \\
  \hline
  $2$ & $2$ & $1$ & $2$ & $3$ & $0$ \\
  \hline
  $3$ & $3$ & $0$ & $2$ & $3$ & $4$ \\
  \hline
  $4$ & $4$ & $1$ & $0$ & $3$ & $4$ \\
  \hline
\end{tabular}

\end{center}
\end{minipage}
\begin{minipage}{0.3\textwidth}
\vspace{0.4cm}
\begin{center}
\input{cayley.table.once.path.1.pic}
\end{center}
\end{minipage}
\begin{minipage}{0.3\textwidth}
\vspace{0.4cm}
\begin{center}
\input{cayley.table.once.path.2.pic}
\end{center}
\end{minipage}
\caption{\textbf{Zig-zagging for words of once-punctured torus
groups}. On the left, the indexed version of Cayley table
\ref{T11CayleyTable}. At the center, the succession of states
related to the word $abA$, that is, $123$ (RL). On the right, the
crash path of word $AabA$.}\label{IndexedCayleyTable}
\end{table}

The Cayley table for once-punctured torus groups is simple to be
implemented in terms of coding: its action just resumes into the
recognition of crash words $\{aA, Aa, bB, Bb\}$. But there are
groups equipped with more complicate presentations, demanding a
generalized management. \emph{The strategy is to take on each
string from the original tree and test its run via the Cayley
table}. We will work on two examples that progressively generalize
concepts.

The once-punctured torus groups offer a comfortable start. The
resulting indexes, originally ranging in $[0-9]$, will be
incremented by $1$ in order to avoid collisions with the crash
state. Let $W=abA$ (RL) and the indexed table
\ref{IndexedCayleyTable} at p. \pageref{IndexedCayleyTable}. $W$
turns into $123$. For algorithmic reasons, the table scan begins
from the `neutral' state of the identity element, at row $0$:
$WI\equiv W$. We read the first symbol $3$ and we move to column
$3$ with state $3$. The sense of a dynamical system is that each
state rules the next one: this explains why we will place at row
$3$. We read the second symbol $2$, we move to column $2$ with
state $2$. We place at row $2$, we read the third symbol $1$ and
we move to column $1$, with (final) regular state $1$.

The second example is the \emph{Klein-four group}, equipped with
simple but less obvious presentation: $\langle a, b |
a^2=b^2=(ab)^2=I\rangle$. Despite of its name, it has nothing to
do with Kleinian groups, but we want to show that the crash state
shall be generalized and no longer tied to pairs of inverse
letters, as we did for once-punctured torus groups: here each
generator is the inverse of itself.
\begin{table}[htb]
\noindent
\begin{minipage}[c]{0.3\textwidth}
    \begin{tabular}{|c|c|c|c|c|}
    \hline
    & $I$ & $a$ & $b$ & $ab$ \\
    \hline
    $I$ & $I$ & $a$ & $b$ & $ab$ \\
    \hline
    $a$ & $a$ & $I$ & $ab$ & $b$ \\
    \hline
    $b$ & $b$ & $ab$ & $I$ & $a$ \\
    \hline
    $ab$ & $ab$ & $b$ & $a$ & $I$ \\
    \hline
    \end{tabular}

\end{minipage}
\begin{minipage}[c]{0.35\textwidth}
\input{klein4grouptree.pic}\label{k4grouptree}
\end{minipage}
\begin{minipage}[c]{0.3\textwidth}
    \begin{tabular}{|c|c|c|c|c|}
    \hline
    & $0$ & $1$ & $2$ & $3$ \\
    \hline
    $0$ & $0$ & $1$ & $2$ & $3$ \\
    \hline
    $1$ & $1$ & $0$ & $3$ & $2$ \\
    \hline
    $2$ & $2$ & $3$ & $0$ & $1$ \\
    \hline
    $3$ & $3$ & $2$ & $1$ & $0$ \\
    \hline
    \end{tabular}
\label{k4grouptreeindexed}
\end{minipage}
\caption{\textbf{Klein Four-Group}. From left to right,
multiplication table, tree model and indexed version of the left
table.}\label{KleinFourTable}
\end{table}

We immediately notice a generator labelled with $ab$, a compound
word. Turning this Cayley table into the indexed version (table
\ref{KleinFourTable}, on the right) will keep up the one-to-one
symbolism. We `expand' a non-cyclic $3$-branched original tree
(like at p. \pageref{tree_full_01}) and apply the mapping
$[a\rightarrow 1, b\rightarrow 2, ab\rightarrow 3]$ to return the
indexed tree. Now we can rework the indexed words via the Cayley
table. Let the indexed $W=123$ (RL), which translates back to
$(a)(b)(ab)$.\footnote{We will explain further why we split it
into tokens.} The reading will end up at row $1$ / column $1$. The
final zero index shows that `123' is a crash word! There are even
more complicate groups, as the one with commutators of order $2$
in \cite{Indras-2002}, p. 359: $(abAB)^2=I$, including the
generators with multiple crash states.

\section{Implementation}\label{implementation}
\emph{Disclaimer}. As higher level language, Javascript runs
reasonably slower than lower level homologous ones, such as
C$^{++}$. The fastest algorithms want bare coding: just the strict
number of calculations and possibly no external calls. Our web
environment\footnote{See footnote \ref{footnote_circle}.} has
broader goals and does not follow such indications; our benchmarks
just attest faster speeds, not the fastest possible. We will show
some Javascript pseudo-code implementing the indexed scan.

We count all nodes in the original tree and return the indexed
word through the routine \texttt{get\_RL\_word}. The goal is to
return strings whose length is equal to the node depth. This
routine does not merely run as the ordinary base conversion. Some
tests showed that there could be base-10 indexes which may not
retrieve the sought word of required depth and then bug the final
rendering. So we put an extra \texttt{if}-statement assuring that
the tree is transversed all the way back to depth $1$. Our code is
tuned to $9$ generators at most, each one binding to one digit
from $1$ to $9$ (and $0$ for $I$). If more than $9$ are required,
we shall replace the string \texttt{out} with a specific array
object that stores indexes as tokens, because a string -- although
being an array too -- allows to save only one symbol at each
position of this data structure.

\vspace{0.4cm}

\begin{Verbatim}[fontsize=\small,frame=single,numbers=left,label=RL-path scanner]
function get_RL_word( _n, _gens_n, _depth )
{
  // _gens_n : is the number of symbols, i.e. of generators
  var _rem = 0, _quot = _n, out = "" ;
  while( true )
  {
    // remainder incremented by 1 to match our indexing
    // rule: 0 for identity, other digits for generators index
    _rem = ( _quot % _gens_n ) + 1 ;
    _quot = ( _quot / _gens_n ) >> 0 ; // integer division
    //it stops only when depth 1 is reached
    if ( _quot < _gens_n && _depth <= 1 ) return _rem + '' + out ;
    out = _rem + '' + out ; // string concatenation
    _depth-- ;
  }
}
\end{Verbatim}

We give the code below to check an indexed word run through any
Cayley table. It is simply a multi-dimensional array reading,
returning $0$ if a crash word is met, otherwise returns
$1$.\vspace{0.4cm}

\begin{Verbatim}[fontsize=\small,frame=single,numbers=left,label=Check word run]
// indexed Cayley table for once-punctured torus groups
var _idx_cayley_table = [ [ 0, 1, 2, 3, 4 ],
                          [ 1, 1, 2, 0, 4 ],
                          [ 2, 1, 2, 3, 0 ],
                          [ 3, 0, 2, 3, 4 ],
                          [ 4, 1, 0, 3, 4 ] ] ;

function check_word_run( _idx_word, _cayley_table )
{

// we start from row 0, by convention
var _idx = -1, _ret = 1, _row = 0 ;

/* get the input word, split indexes into tokens, convert'em all
into numbers and reverse for RL order */

_idx_word=_idx_word.split( "" ) ;
_idx_word=_idx_word.map(function(_i){return parseInt(_i,10 );}) ;
_idx_word = _idx_word.reverse(); //RL reading order

for( var _i = 0 ; _i < _idx_word.length ; _i++ )
{
  _idx = _idx_word[_i], _row = _idx_cayley_table[_row][_idx] ;
  if ( _row == 0 ) { _ret = 0; break ; } // crash state is met
}

return _idx == -1 ? 0 : _ret ;
}
\end{Verbatim}

All nodes will be scanned according to the tree in fig.
\ref{tree_indexed_pic} and keep track of both depth and run for
each node. The attractor can be plot in `\emph{limit set}' or
`\emph{tiling}' rendering mode, whether word points are drawn for
leaves only or for all nodes respectively. Given an $m$-branched
tree, let $\lambda=m^d$ be the number of nodes at bounded depth
$d\leq D <+\infty$. The tree counts $\sigma=\sum_{d=1}^{D}\lambda$
or $\lambda=m^D$ nodes if the rendering works in `tiling' or in
`limit set' mode respectively. We finally code the nested loops
below, to render the attractor in four steps: (1) a loop feeding
the fixed points to start the orbits;\footnote{See table in
\cite{Indras-2002}, p. 135.} (2) a loop to get the \emph{indexed
word} from each node; (3) a loop to read and compute and draw the
word values.

\vspace{0.4cm}

\begin{Verbatim}[fontsize=\small,frame=single,numbers=left,label=Index search algorithm]
var _gens_n = 4, _rl_word = "", _max_depth = 4, _fp = null ;
/*assume that we already collected the fixed points of the Mobius
  transformations of K into the array _input_fixed_pts and that we have
  a multi-dimensional array storing the current Cayley Table*/
// feed fixed points
for( var _p = 0 ; _p < _input_fixed_pts.length ; _p++ )
{
    _nodes_n = Math.pow( _gens_n, _d ) ;
    // n is the index of each node belonging to depth _d
    for( var _n = 0 ; _n < _nodes_n ; _n++ )
    {
      _fp = _input_fixed_pts[_p] ;
      _rl_word = get_RL_word( _n, _gens_n, _d ) ;
      // pseudo-code (loop):
      // 1.0 read _rl_word
      // 2.0 check_word_run( _rl_word, _cayley_table )
      // 2.1 if crash state is met, continue to the next iteration
      // 2.2 otherwise, for each digit in the _rl_word:
      // 2.2.1 get the related Mobius transformation M_n.
      // 2.2.2 apply _fp = M_n(_fp)
      // 3.0 draw _fp on the screen, according to
             'tiling' (any word)
             'limit set' (only words whose length = depth) mode
    }
}
\end{Verbatim}

\section{Conclusions}\label{conclusions}
The benefits of this new version of the deterministic approach
account to: 1) pull out any word from an integer; 2) save memory
resources required by the dictionary array and by calls to outer
functions for trasversing the pruned tree;\footnote{See
breadth-first and depth-first implementations, p. 115 and 148--151
of \cite{Indras-2002} respectively.} 3) very easy implementation.

The deterministic approach, as well as the probabilistic version,
runs too slow long to get fine results at reasonable times: both
of them do need some tricks to be accelerated.

\bibliographystyle{amsplain}

\end{document}